\documentclass[aps,pra,nofootinbib,floatfix,preprintnumbers,tightenlines,11pt,superscriptaddress]{revtex4-2}
\pdfoutput=1

\usepackage{amsmath,amssymb,amsfonts,amsthm}
\usepackage{verbatim}
\usepackage{algorithmic}
\usepackage{theoremref}
\usepackage{mathrsfs}
\usepackage{algorithm2e}
\RequirePackage[OT1]{fontenc}
\RequirePackage{hyperref}

\usepackage{bm}
\usepackage[dvipsnames]{xcolor}
\usepackage{titlesec}
\usepackage{textcomp}
\usepackage{fullpage}
\usepackage{bbm}
\usepackage[section]{placeins}
\usepackage{comment}

\usepackage{graphicx}
\usepackage{subfigure}
\usepackage{caption}
\usepackage{float}
\graphicspath{ {./figures/} }

\def\BibTeX{{\rm B\kern-.05em{\sc i\kern-.025em b}\kern-.08em
    T\kern-.1667em\lower.7ex\hbox{E}\kern-.125emX}}

\usepackage{hyperref}
\hypersetup{
     colorlinks   = true,
     citecolor    = blue
}

\usepackage{math-macros}

\begin{document}

\title{Quantum  Persistent Homology for Time Series}

\author{Bernardo Ameneyro}
\email{bameneyr@vols.utk.edu}
\affiliation{Department of Mathematics, The University of Tennessee, Knoxville, TN 37996-1320, USA}
\author{George Siopsis}
\email{siopsis@tennessee.edu}
\affiliation{Department of Physics and Astronomy, The University of Tennessee, Knoxville, TN 37996-1200, USA}
\author{Vasileios Maroulas}
\email{vasileios.maroulas@utk.edu}
\affiliation{Department of Mathematics, The University of Tennessee, Knoxville, TN 37996-1320, USA}
\date{\today}

\begin{abstract}
Persistent homology, a powerful mathematical tool for data analysis, summarizes the shape of data through tracking topological features across changes in different scales.
Classical algorithms for persistent homology are often constrained by running times and memory requirements that grow exponentially on the number of data points.
To surpass this problem,  two quantum algorithms of persistent homology have been developed based on two different approaches. However, both of these quantum algorithms consider a data set in the form of a point cloud, which can be restrictive considering that many data sets come in the form of time series. In this paper, we alleviate this issue by establishing a quantum  Takens's delay embedding algorithm, which turns a time series into a point cloud by considering a pertinent embedding into a higher dimensional space. Having this quantum transformation of time series to point clouds, then one may use a quantum persistent homology algorithm to extract the topological features from the point cloud associated with the original times series.
\end{abstract}

\maketitle


\section{Introduction} \label{sec:intro}
Topological data analysis (TDA) methods capture shape properties of data, and they have found a  number of applications from biology \cite{tda_wheeze,tda_clustering2015,gunnarcancer, HodgeCycle,CPD_me, Maroulas2021} to chemistry and materials science \cite{Maroulas2020,Townsend2020,NA2022, Papamarkou2022,Chen2021}, and from classification and clustering for action recognition \cite{tda_action}, and handwriting analysis \cite{tda_number}, to classification and clustering of signals \cite{MaMa16,Marchese2018,tda_action,tda_timeseries,tda_wheeze}.
Persistent homology is a powerful technique, on which TDA relies, and summarizes nonlinear and high-dimensional data retaining useful information about its shape.

The topological features discovered via persistent homology are displayed in a persistence diagram, and the efficient computation of such diagrams using packages such as Dionysus \cite{dionysus} and Ripser \cite{ripser} leverage certain properties of simplicial complexes to create persistence diagrams efficiently \cite{persistwist}. However, these classical algorithms  are not easily scalable and tend to deteriorate with the increase in the number of data size.
Indeed, a point cloud consisting of \(n\) points possesses \(2^n\) potential subsets that could contribute to the topology, so the best classical algorithm for estimating these features with accuracy \(\delta\) takes time \(O(2^n \log (1/\delta))\) and requires \(O(2^{n})\) bits to encode the subsets \cite{zomo-tda}.

To bypass the computational bottleneck, quantum algorithms for homology (fixed data resolution/scale) \cite{lloyd2016quantum, siopsis2019quantum}, and more recently, for persistent homology (varying data resolutions/scales) \cite{ameneyro2022quantum, hayakawa2021quantum} use a quantum random access memory (QRAM) to efficiently read a point cloud data set, and encode it into pertinent quantum states.
One then uses these quantum states to build membership oracles that identify the features that are present at different scales.
The main advantage of these algorithms is that they use relatively small QRAMs of size \(O(n^{2})\), which only require \(O(\log n)\) calls to access the data \cite{lloyd2008qram}, as opposed to classical algorithms that need memories of size \((O(2^{2n})\).

These quantum algorithms make use of membership oracles to determine which objects are present at a certain scale.
The oracles are simply subroutines that take into consideration a pertinent triangulation of point cloud data, called simplex and a positive number \(\eps\) encoded into qubits and return a qubit in state \(|1\ra\) if the simplex is present at scale \(\eps\), or a qubit in state \(|0\ra\) otherwise.
On the other hand, the membership oracles that exist in the literature so far are merely restricted to simplicial complexes built from point cloud data sets.

In this paper, a quantum TDA algorithm is proposed by providing a membership oracle for time series.
Indeed, a QRAM is queried to access the time series, and a quantum version of the Takens's delay embedding theorem \cite{takens} is established for transforming a time series into a point cloud while preserving the relevant topological information.
In turn, the oracle is plugged into the quantum algorithm for persistent homology \cite{ameneyro2022quantum} to study time series and their behavior (periodic, chaotic, etc.) by analyzing their corresponding point cloud. 

This paper is organized as follows. Section \ref{sec:background} introduces the background information like the Takens's delay embedding, and summarizes classical and quantum algorithms for TDA. Section \ref{sec:methods} details our novel membership oracle as well as the quantum TDA algorithm for time series.
Section \ref{sec:results} shows two implementations of the quantum algorithm to time series data sets, and
finally, Section \ref{sec:conclusion} concludes with a brief discussion of our results.

\section{Background}\label{sec:background}
\subsection{Delay Embedding}

Consider a time series \(x_{t}\) for \(t=1,2,...,T\). The Takens's delay embedding of the time series $x_t$ is given by the vector

\begin{equation} \label{eq:embedding}
    v_{i} = (x_{i}, x_{i + \tau}, \dots , x_{i + (d-1)\tau})
\end{equation}

\noindent for \(i = 1,...,T -  \tau d\), where \(d\) is the dimension of the point cloud, and \(\tau\) is the delay parameter.
The Takens's delay embedding theorem guarantees that the map from the time series to point cloud is an embedding and as such all the relevant topological information is retained \cite{takens}.
The parametric choices of the dimension, $d$, and the delay, $\tau,$ has been widely discussed in the literature, e.g. see \cite{Marchese2016} and references therein. For example, the time series is typically embedded into a  $d=2, 3,$ space, and the delay parameter may be selected based on the autocorrelation.  

\subsection{Vietoris-Rips Filtration and Persistent Homology}

\begin{figure}[btp]
    \centering
    \includegraphics[clip, trim = {0 0 0 10}, width=0.6\linewidth]{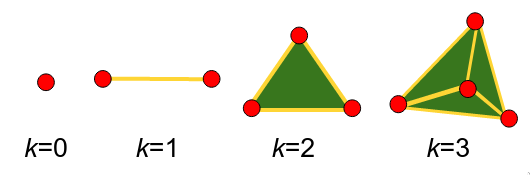}
    \captionof{figure}{Simplices of dimensions \(k = 0, 1, 2, 3\).}
    \label{fig:simp}
\end{figure}

An oriented \(k-\)simplex or simplex of dimension \(k\) is the collection of all convex combinations formed by \((k+1)\) ordered and linearly independent vertices, a few examples are shown in Fig. \ref{fig:simp} such as vertices, edges, triangles, tetrahedrons.
They are written as the ordered list of their vertices, and a subset of this list is called a face of the simplex.

Homology studies simplicial complexes and chain complexes.
A simplicial complex is a collection of simplices such that the faces of any simplex in the complex are also in the complex and the intersection of any two simplices in the complex is also a simplex in the complex. A commonly used simplicial complex is the Vietoris-Rips (VR) complex, which at scale \(\eps\) is defined by all the simplices with a diameter of at most \(\eps\).
So, a simplex \(\s\) is in the complex \(S^{\eps}\) if and only if all the pairwise distances between its points are less than \(\eps\).
On the other hand, the chain complex of \(S^{\eps}\) over \(Z_{3} = \{-1, 0, 1\}\) is the collection of formal sums of simplices in \(S^{\eps}\) with coefficients in \(Z_{3}\).

Classical algorithms encode the VR complex and chain complex, then compute its corresponding boundary matrix, and finally diagonalize it to extract topological features like the number of connected components, holes, voids, and \(k\) dimensional holes in general.

Algorithms for persistent homology consider instead a filtration or nested sequence of VR complexes \(S^{\eps_{0}} \subseteq \dots \subseteq S^{\eps_{N}}\) and analyze a boundary matrix in which the simplices are ordered according to the sequence \cite{zomo-tda}.
Persistence diagrams are used to display the results of these algorithms.
Points on the persistence diagram \((b, d)\) represent the number of \(k\) dimensional holes born at scale \(b\) that disappear at scale \(d\).

\subsection{Quantum TDA Algorithms for point cloud data}

The quantum algorithms appeared in \cite{ameneyro2022quantum, hayakawa2021quantum} established for the first time a quantum implementation for computing the persistent homology of a point cloud to reveal its topological properties across different scales. The study in \cite{ameneyro2022quantum} defines a linear map, the persistent Dirac operator, which tracks topological features across a pair of different scales.
The persistent Dirac operator is a generalization of the Dirac operator established in the algorithms of  \cite{lloyd2016quantum, siopsis2019quantum}, which cannot track features across different scales/resolutions, and therefore they cannot obtain any persistent information.
The work in  \cite{hayakawa2021quantum} builds a block-encoding of the persistent combinatorial Laplacian and uses the quantum singular value transformation to extract the persistent topological features.

All of these quantum algorithms use a QRAM to store and access the data efficiently, they encode the points of a cloud as qubits and use superpositions to represent complexes.
They also rely on a membership oracle to identify the simplices that are present at a certain scale and construct projections onto the corresponding VR complexes.

\section{Quantum Methodology for Time Series} \label{sec:methods}
\subsection{Encoding Data and Simplices}

Consider a discrete time series \(x_t\),  \(t=1, 2 \dots, T\),  such that $T >0$ is an  integer.
The data are stored in a QRAM, allowing for their quantum parallel access.
The QRAM maps a quantum state \(|t\ra |0\ra\) to the state \(|t\ra |x_t\ra\) where \(|x_t\ra\) is a state normalized to reflect the value \(x_t\). Such states can be encoded using \(O(\log_2T)\) qubits.

Recall that the point cloud relates to the time series via the vectors $v_{i}$ given in Eq. \eqref{eq:embedding}.
We use \(T - \tau d \) qubits to encode these points so that any simplex \(\s\) formed with these points is represented by a quantum state \(|\s\ra\), with the qubits corresponding to points in the simplex in state \(|1\ra\), and all other qubits in state \(|0\ra\).
A balanced superposition of the states represents the maximal simplicial complex \(S\). Define \(\Z_{3} = \{-1, 0, 1\}\).
Using these quantum states as a basis of the generated Hilbert space over $\Z_3$ encodes the chain complex needed to extract the desired topological features.

\subsection{Quantum Delay Embedding}
Considering the time series data \(x_{t}\), along with the embedding parameters, the dimension, \(d\), of the space to which  the times series is embedded, and the delay parameter, \(\tau\), one may check for membership in a simplicial complex by first computing the distances between points $v_i$ in the embedded point cloud, defined in Eq. \eqref{eq:embedding}.
The distance between two points \(v_i\) and \(v_j\) of the delay embedded time series is given by

\begin{equation} \label{eq:distance}
  D(v_i, v_j) = \max_{0 \le t < d} |x_{i + t\tau} - x_{j + t\tau} |.
\end{equation}

With the ability to create quantum states encoding the values \(x_t\), we can construct a quantum circuit that takes input states of the form \(|t_1\ra |t_2\ra |0\ra\) and returns the output state \(|t_1\ra |t_2\ra \left| |x_{t_1} - x_{t_2}|^2 \right\ra\), where the last register contains an estimate of the difference between \(x_{t_1}\) and \(x_{t_2}\).
To obtain an estimate with accuracy \(\delta\) it is necessary to make \(O(\delta^{-1})\) calls to the QRAM and \(O\left(\delta^{-1} (\log_2T)^2\right)\) \cite{lloyd2008qram, lloyd2016quantum}.
Moreover, this circuit can also operate in quantum parallel.

Notice that a simplex \(\s\) is in the VR complex \(S^{\eps}\) if and only if for each pair of vertices \(v_i\) and \(v_j\) in the simplex, $D(v_i,v_j) \leq \eps,$ where the distance, $D,$ is defined in Eq. \eqref{eq:distance}. 
Being able to compute the difference \(|x_{t_1} - x_{t_2}|\) allows us to create an oracle \(\mathcal{O}^{\eps}\) that verifies if the difference \(|x_{t_1} - x_{t_2}| \leq  \eps\). 
Such an oracle takes as input the state \(|t_1\ra |t_2\ra |1\ra\) and returns the state \(|t_1\ra |t_2\ra |a^{\eps}\ra\), where

\begin{equation}
  a^{\eps} = \left\{
    \begin{array}{ccc}
      1 & , & |x_{t_1} - x_{t_2} | \le \eps \\
      0 & , & |x_{t_1} - x_{t_2} | > \eps
    \end{array}
  \right.
\end{equation}

Given a simplex \(\s\) encoded as the quantum state \(|\s\ra\), where the qubits in state \(|1\ra\) correspond to the vertices of \(\s\), we can make repeated calls to the oracle \(\mathcal{O}^{\eps}\) to check if the simplex belongs to the VR complex at scale \(\eps\).
In particular, we need to verify the inequality for all \(t_1 = i + t\tau\) and \(t_2 = j + t\tau\) with \(0 \le t < d\) and \(i < j\) such that \(v_i\) and \(v_j\) are vertices of \(\s\), where $d$ and $\tau$ as in Eq. \eqref{eq:embedding} must be chosen according to the data.
The vertices \(v_i, v_j\) are obtained from the quantum state \(|\s\ra\), and they correspond to the qubits in state \(|1\ra\).
One concludes that \(\s\) is in \(S^{\eps}\) if all the calls to the oracle \(\mathcal{O}^{\eps}\) yield an output with the last qubit in state \(|1\ra\).
However, if at least one of the calls returns an output with the last qubit in state \(|0\ra\), it means the simplex is not yet present at that scale.

To that end, for fixed $d, \mbox{and} \;\tau ,$ a membership oracle \(\mathcal{O}^{\eps}_{d, \tau}\) is constructed such that it acts on a quantum state \(|\s\ra |1\ra\) according to

\begin{equation} \label{eq:membership_oracle}
  \mathcal{O}_{d, \tau}^\eps |\s\ra |1\ra = \left\{
    \begin{array}{ccc}
      |\s\ra |1\ra & , & \s \in S^{\eps} \\
      |\s\ra |0\ra & , & \s \notin S^{\eps}
    \end{array}
  \right. .
\end{equation}

The membership oracle, \(\mathcal{O}^{\eps}_{d, \tau}\), makes at most \(d k (k+1) / 2\) calls to the oracle \(\mathcal{O}^{\eps}\) in order to determine whether a simplex \(\s\) of dimension \(k\) is present in the VR complex at scale \(\eps\).

\subsection{Quantum Algorithm for Persistent Homology}

Having established a membership oracle \(\mathcal{O}^{\eps}_{d, \tau}\), defined in Eq. \eqref{eq:membership_oracle}, allows to consider a quantum persistent homology algorithm for time series. First, one uses \(T - \tau d\) qubits to represent the point cloud points, $v_i$, defined in Eq. \eqref{eq:embedding}, which result from the embedding.
Then, a simplex \(\s\) is represented by the quantum state \(|\s\ra\), where the qubits corresponding to the vertices of \(\s\) are in state \(|1\ra\), and all others are in state \(|0\ra\). The boundary map \(\partial\) is essential to extract the topological features and it can be encoded using Pauli \(X\) operators, which are switches that change state \(|0\ra\) into state \(|1\ra\) and vice versa.

The operator given in Eq. \eqref{eq:boundary} maps a simplex \(|\s_k\ra\) of dimension \(k\) into a superposition of the \((k-1)\)-simplices that conform its boundary, which are none other than the simplices obtained by removing each of the vertices in \(\s_k\),

\begin{equation}\label{eq:boundary}
  \partial_k |\s_k\ra = \sum_{l=0}^{k} (-1)^l X_{i_l} |\s_k\ra.
\end{equation}

Given two fixed scales \(\eps \le \eps'\), the membership oracle in Eq. \eqref{eq:membership_oracle} is used to implement the projection operators, \(P^{\eps}\) and \(P^{\eps'}\), onto the subspaces that encode the VR complexes \(S^{\eps}\) and \(S^{\eps'}\). The boundary operator and the projections are in turn used to build a persistent Dirac operator

\begin{equation}
  \label{eq:dirac}
  B_{k}^{\eps, \eps'} = \begin{pmatrix}
     -\xi I & \partial_{k}^{\eps, \eps} & 0 \\
     \partial_{k}^{\eps, \eps\dagger} & \xi I & \partial_{k+1}^{\eps, \eps'} \\
     0 & \partial_{k+1}^{\eps, \eps'\dagger} &  -\xi I 
  \end{pmatrix} ,
\end{equation}

\noindent where \(\xi\) is an integer that is chosen according to the particular data at hand, and \(\partial_{k+1}^{\eps, \eps'}\) is used to denote \(P^{\eps}\partial_{k+1}P^{\eps'}\).

The parameter \(\xi\) in Eq. \eqref{eq:dirac} is an eigenvalue of the persistent Dirac operator and its eigenspace corresponds to the kernel of a generalization of the graph Laplacian \cite{ameneyro2022quantum}.
In particular, the multiplicity of this eigenvalue is the persistent Betti number \(\beta_{k}^{\eps, \eps'}\), which counts the number of \(k\)-dimensional holes in the point cloud that persist from scale \(\eps\) up to at least scale \(\eps'\).

To estimate a persistent Betti number, the quantum phase estimation algorithm is performed on the persistent Dirac operator as in \cite{ameneyro2022quantum} to derive the following probability distribution

\begin{equation}
  \label{eq:phase}
  \HP (p) = \dfrac{1}{N} \sum_{\lambda_s} \dfrac{1}{M} \dfrac{\sin^2 \pi l \lambda_s}{\sin^2\frac{\pi(l\lambda_s - p)}{M}},
\end{equation}

\noindent where \(\lambda_s\) are the eigenvalues of the persistent Dirac operator in Eq. \eqref{eq:dirac} and \(N\) its dimension, while \(l\) and \(M\) are parameters to be chosen according to the data. 

Next, one may use the Eq. \eqref{eq:phase} directly to estimate the persistent Betti numbers. Notice that if \(l\) and \(M\) are large enough, each of the terms in the sum in Eq. \eqref{eq:phase} is either equal to 1, if \(p = l\lambda_s\), or very close to 0, if  otherwise. Therefore \(\HP(p)\) of Eq. \eqref{eq:phase} is proportional to the multiplicity of \(\lambda_s\) at \(p = l\lambda_s\) and 0 everywhere else.
The desired persistent Betti number is then given by
\begin{equation} \label{eq:pers_betti}
\beta_{k}^{\eps, \eps'} = N \HP (l\xi), 
\end{equation}
where the parameters \(l\) and \(M\) were chosen according to the eigenvalues of the persistent Dirac operators from Eq. \eqref{eq:dirac}. Indeed, 
the parameter \(l\) was picked as the inverse of the gap between \(\xi\) and the closest eigenvalue, then \(M\) was chosen as a power of 2 larger than the greatest eigenvalue (in norm) of the persistent Dirac operator times \(l\).

In order to obtain all persistent Betti numbers at various scales and represent them onto  persistence diagrams, one needs to consider an increasing sequence of scales \(\eps_0 < ... < \eps_n\), chosen according to the data.
Then one implements the aforementioned  quantum methodology to obtain the persistent Betti numbers for each pair \(\eps_i, \eps_j\) with \(0 \le i \le j \le n\).
Since the estimations are independent, this can be done in parallel.
Finally, once we have all possible Betti numbers, and the scales, which appear and disappear, one may depict them onto persistence diagrams.
Indeed, if \(\mu_{k}^{\eps, \eps'}\) denotes the number of \(k-\)dimensional holes that appear in scale \(\eps\) and disappear at scale \(\eps'\), we have
\(  \mu_{k}^{\eps_i, \eps_j} = \beta_{k}^{\eps_i, \eps_{j-1}} - \beta_{k}^{\eps_i, \eps_{j}} - \left( \beta_{k}^{\eps_{i-1}, \eps_{j-1}} - \beta_{k}^{\eps_{i-1}, \eps_{j}} \right)\), where $\beta_k$ is defined in Eq. \eqref{eq:pers_betti} for  various scales.

\section{Results} \label{sec:results}
Two time series are considered to discover their shape properties. The first data is a simulated periodic discrete time series, and the second is a segment from an electroencephalography (EEG) measurement of a brain taken while the subject listened to music \cite{eeg-data}. In both cases the task is to  employ the quantum framework of Section \ref{sec:methods} by first embedding the time series into a point cloud using the quantum delay embedding, and then examining the topological features of the corresponding embedded point clouds using the quantum persistent homology algorithm.   
 
\subsection{Periodic time series example}
\begin{figure}[btp]
    \centering
    \includegraphics[width=0.75\linewidth]{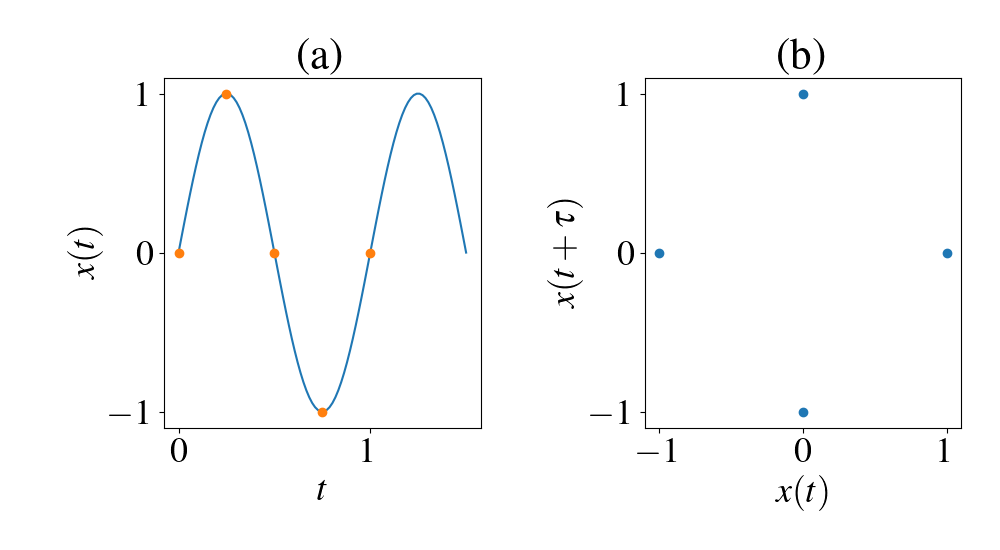}
    \captionof{figure}{(a) The graph of \(\sin(2\pi t)\) (blue line) along with a discrete time series (orange dots) given by \(t = 0, 1/4, 1/2, 3/4, 1\). (b) The point cloud obtained by Takens's delay embedding using \(\tau = 1\) and \(d=2\).}
    \label{fig:time-series-one}
\end{figure}

Consider the discrete time series based on the periodic function \( \sin(2\pi t) \). Using a delay \(d=2, \; \tau = 1\), the  time series  is embedded into the 2-dimensional point cloud as shown in Fig. \ref{fig:time-series-one}.
Notice that this function has one maximum (or minimum) per period, which  translates to one hole in the corresponding embedded point cloud, constructed  by the quantum Taken's delay embedding.

\begin{figure}[btp]
    \centering
    \includegraphics[width=0.75\linewidth]{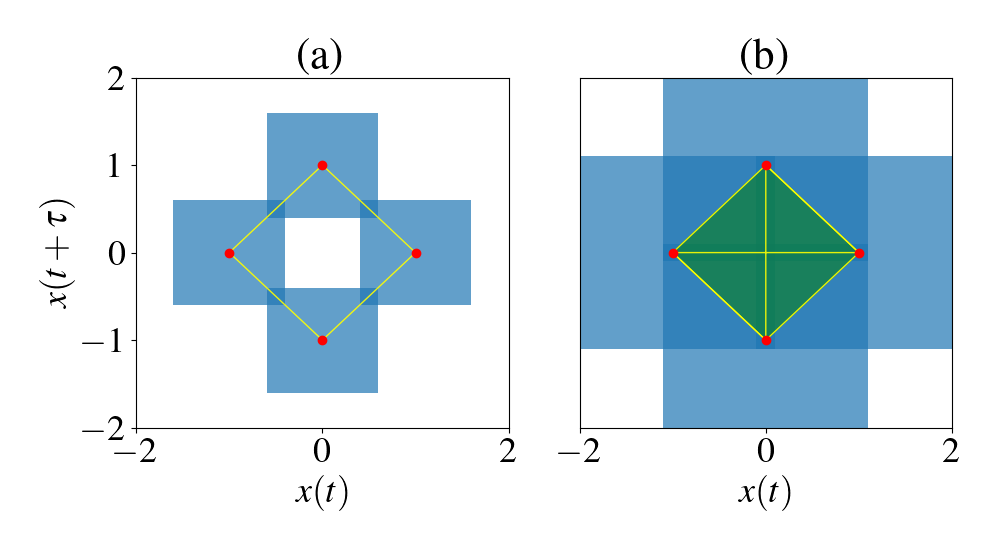}
    \captionof{figure}{Vietoris Rips complexes of the embedded point cloud associated with the sinusodial signal of Fig. \ref{fig:time-series-one} at scales (a) \(\eps_1 = 1.2\), the points are pairwise connected, which results in a hole in the middle;  and (b) \(\eps_2 = 2.2\), the points become totally connected and the hole disappears. The blue squares are the balls of diameter \(\eps_{i}\) around each point. }
    \label{fig:simplices}
\end{figure}

\begin{figure}[btp]
    \centering
    \includegraphics[width=0.45\linewidth]{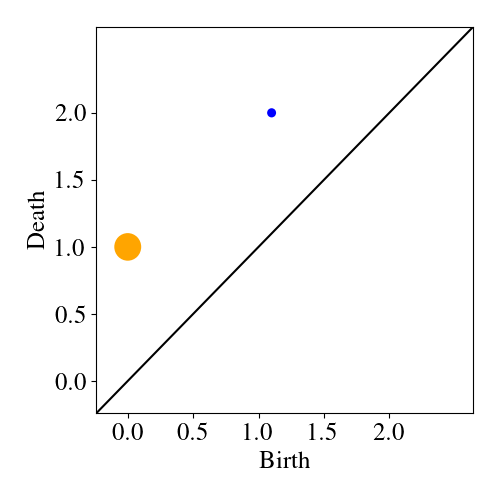}
    \captionof{figure}{Persistence diagram for the time series in Fig \ref{fig:time-series-one}. The horizontal axis marks the scales at which the topological features are born, while the vertical axis marks the scales at which they disappear.
    The size of the dots represents the number of features that appear and disappear at the same scales.
    Finally, the orange dots along the vertical axis represent the connected components, while the blue dots closer to the diagonal line are the one dimensional holes.}
    \label{fig:diagram-one}
\end{figure}

To construct the persistence diagram, an increasing sequence of twenty-five different scales starting at \(\eps_0 = 0.0< \eps_1 = 0.1< \dots < \eps_{24} = 2.4\) with step size 0.1 taken into account.
Using the quantum persistent homology algorithm, the persistent Betti numbers of dimensions 0 and 1 for each possible pair of scales are computed, and depicted in Fig. \ref{fig:diagram-one}.
As expected, a single one dimensional hole exists due to periodicity, and this is born at scale 1 and died at scale 2.
On the other hand, the number of connected components, which is initially four, is reduced to one when the one dimensional hole is formed at scale 1.

\subsection{An electroencephalogram example}
\begin{figure}[btp]
    \centering
    \includegraphics[clip, trim = {0 10 0 0},width=0.75\linewidth]{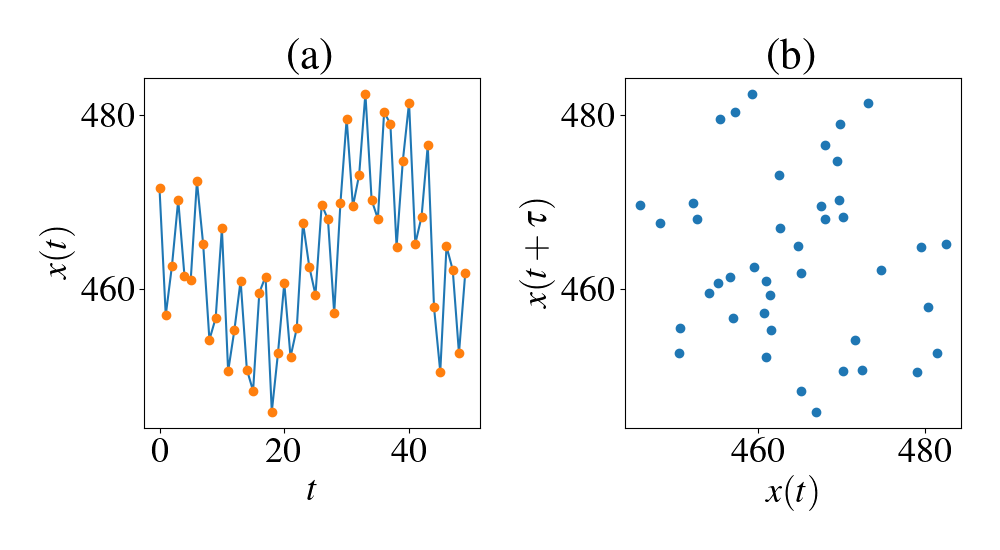}
    \captionof{figure}{(a) A segment of an EEG signal measured while the subject listened to music \cite{eeg-data}. (b) The point cloud obtained by the Takens's delay embedding using \(\tau = 8\) and \(d=2\).}
    \label{fig:time-series-eeg}
\end{figure}

 A segment of fifty  measurements from an electroencephalogram (EEG) signal  while the subject listened to music is considered in Fig. \ref{fig:time-series-eeg}. 

\begin{figure}[btp]
    \centering
    \includegraphics[clip, trim = {0 0 0 20}, width=0.45\linewidth]{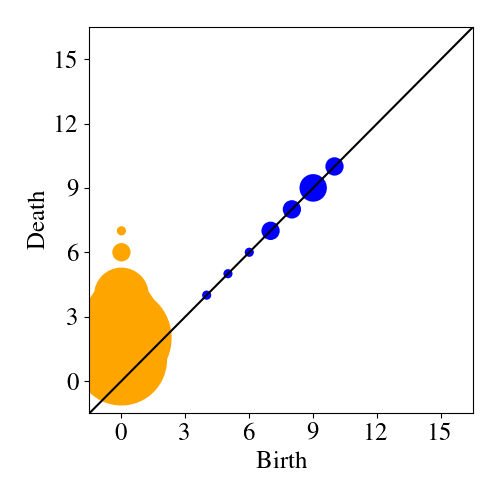}
    \captionof{figure}{Persistence diagram for the time series in Fig \ref{fig:time-series-eeg}. The horizontal axis states the scales at which the topological features are born, while the vertical axis marks the scales at which they disappear.
    The size of the dots represents the number of features that appear and disappear at the same scales.
    Finally, the orange dots along the vertical axis represent the connected components, while the blue dots closer to the diagonal line are the one dimensional holes.}
    \label{fig:diagram-eeg}
\end{figure}

The point cloud obtained using the quantum delay embedding algorithm for a delay \(\tau = 8\) and \(d = 2\) is shown in Fig. \ref{fig:time-series-eeg}. Next, the persistence diagrams are constructed using the quantum persistent homology algorithm for an increasing sequence of scales $\eps_k = k,$ for $k = 0, \dots, 15$.
The persistence diagram of dimensions 0 (connected components) and 1 (holes) are depicted in Fig. \ref{fig:diagram-eeg}.
In this case one may see a number of small one dimensional holes that appear and disappear very quickly indicating that the segment of the signal is not periodic.
The number of connected components is reduced quickly as shown by the large circles near the origin.
Nevertheless, a few survive past scale 6 which indicates the presence of clusters in Fig. \ref{fig:time-series-eeg} (b).

\section{Conclusion and Discussion} \label{sec:conclusion}
This paper makes possible for the first time, the topological data analysis of a time series by providing a quantum delay embedding of a time series into a point cloud. 
We introduced a new membership oracle for time series data, by changing the way in which the data is stored in the QRAM, and how it is used in the quantum topological data analysis algorithms to construct the relevant linear operators.
The membership oracle provided here is for Vietoris-Rips complexes, but it is possible that a similar approach could be extended for other types of complexes dependent on the  the data. 
For example, lazy witness complexes are similar to VR complexes and retain the same topological characteristics, but in order to check for membership it suffices to consider only a subset of the vertices.
This could further reduce the amount of resources that the quantum algorithm requires to extract the persistence information, and we plan to examine this in the future.

\acknowledgments

This work has been partially supported by the NSF  DMS-2012609, and DGE-2152168.

\bibliographystyle{unsrt}
\bibliography{references}

\end{document}